\documentclass[twocolumn,amsmath,amssymb,prd]{revtex4}

\usepackage{wasysym}

\def\al{\alpha}
\def\be{\beta}
\def\ga{\gamma}

\def\ep{\epsilon}

\def\et{\eta}

\def\rh{\rho}

\def\si{\sigma}

\def\ch{\chi}
\def\ps{\psi}
\def\om{\omega}

\def\Th{\Theta}

\def\Om{\Omega}
\def\mn{{\mu\nu}}

\def\cL{{\cal L}}

\def\fr#1#2{{{#1} \over {#2}}}
\def\half{{\textstyle{1\over 2}}}
\def\quar{{\textstyle{1\over 4}}}

\def\frac#1#2{{\textstyle{{#1}\over {#2}}}}

\def\prt{\partial}

\def\etal{{\it et al.}}

\def\pt#1{\phantom{#1}}

\def\ab{\overline{a}{}}
\def\bb{\overline{b}{}}

\def\db{\overline{d}{}}
\def\eb{\overline{e}{}}

\def\gb{\overline{g}{}}
\def\Hb{\overline{H}{}}

\def\sb{\overline{s}{}}

\def\afp{(a^{\prime{\rm B}}_{\rm{eff}})}

\def\afb{(\ab_{\rm{eff}})}

\def\abs{(\ab^{\rm S}_{\rm{eff}})}

\def\abE{(\ab^{\oplus}_{\rm{eff}})}

\def\abn{(\ab^n)}

\def\ebn{(\eb^n)}

\def\afw{(a^w_{\rm{eff}})}
\def\afe{(a^e_{\rm{eff}})}
\def\afp{(a^p_{\rm{eff}})}
\def\afn{(a^n_{\rm{eff}})}

\def\bwt{\widetilde{b}{}}

\def\bgm{(b_{\rm gm})}
\def\bgmas{(b_{{\rm gm,} a, s})}
\def\bgmE{(b_{{\rm gm,} \oplus})}
\def\bgmS{(b_{{\rm gm,} \odot})}

\def\bnif{(b_{\rm nf})}

\newcommand{\beq}{\begin{equation}}
\newcommand{\eeq}{\end{equation}}
\newcommand{\bea}{\begin{eqnarray}}
\newcommand{\eea}{\end{eqnarray}}
\newcommand{\bit}{\begin{itemize}}
\newcommand{\eit}{\end{itemize}}
\newcommand{\rf}[1]{(\ref{#1})}

\begin{document}

\title{Lorentz violation, gravitomagnetism, and intrinsic spin}

\author{Jay D.\ Tasson}

\affiliation{Physics Department,
Carleton College,
Northfield, MN 55057}

\date{August 2012}

\begin{abstract}

A largely unconstrained set
of relativity-violating effects
is studied via the gravitomagnetic effect
on intrinsic spins.
The results of existing comagnetometer experiments
are used to place constraints
on two new combinations of these effects
at the 10\% level.
We show that planned improvements in these experiments will make
them competitive with the best existing sensitivities
to this elusive class of relativity-violating effects.
Prospects for measuring the conventional General-Relativistic
gravitomagnetic effect
are also considered.

\end{abstract}

\maketitle

\section{Introduction}
\label{intro}
Our present understanding of nature
at the most fundamental level
relies heavily on Einstein's theories
of Special and General Relativity.
Lorentz symmetry,
the invariance of the laws of physics
under rotations and boosts,
is a foundational assumption
of both of these theories.
Testing such fundamental symmetries
strengthens the experimental foundation
of existing theories
and offers the opportunity to detect
hints of the elusive quantum-consistent theory
at the Planck scale \cite{ksp}.

The gravitational Standard-Model Extension (SME)
is a comprehensive theoretical framework
for performing Lorentz-violation searches \cite{ck,akgrav}.
The framework was constructed
by adding all possible Lorentz-violating terms
to know physics as described by the actions
of the Standard Model of particle physics
and General relativity.
The Lorentz-violating terms
involve coefficients for Lorentz violation,
which can be measured or constrained experimentally. 
A large number of experimental results
have been obtained in the context of the SME \cite{data}.
One class of experiments
that has achieved impressive sensitivity has sought
anomalous precessions of 
intrinsic spins in flat spacetime.
Such experiments will be referred to as anomalous spin-precession experiments
throughout this work.
This class of experiments involves tests with macroscopic spin-polarized solids \cite{bksolid}
and certain tests that can be thought of as clock-comparison tests \cite{clclock}
in which the frequencies depend on spin.
As the level of sensitivity achievable in these experiments
has evolved,
they have become sensitive quantum gyroscopes \cite{bh,mr},
detecting the fact that they are in a rotating reference frame
attached to the Earth,
and they have been reinterpreted as searches for spacetime torsion,
placing the best constraints on constant background torsion \cite{torsion}.
These interpretations exploit the fact that these phenomena
have the same coupling to intrinsic spin
as the SME coefficients for Lorentz violation
originally sought.

Gravitomagnetism is another effect
with this same spin coupling \cite{lvgap}.
As its name suggests,
gravitomagnetism is a gravitational effect arising in analogy
with classical electrodynamics \cite{ht}.
Though the full theory of General Relativity
is highly nonlinear,
it is well known
that the leading gravitational effects
due to weak fields and slow-moving matter
appear as analogues of the electric and magnetic fields
of Maxwell electrodynamics.
The relevant fields are known as the gravitoelectric and gravitomagnetic fields,
and they
generate an analogue of a Lorentz force on a moving test mass \cite{lens}.
The analogy continues to spin precession,
with gravitational fields generating
a precession of angular momenta
just as electromagnetic fields generate
a precession of magnetic moments \cite{apap,lht}.
This analogy has already been extended to the case of 
Lorentz violation
in the pure-gravity sector of the SME \cite{qbgmag},
and the gravitomagnetic precession
of classical angular momenta has now been observed \cite{gpb}.

In this work,
we show that anomalous spin-precession experiments
now have sensitivity to Lorentz-violating contributions to gravitomagnetism,
which arise at lower post-newtonian order
than the conventional gravitomagnetic effects of General Relativity.
Two constraints on presently unconstrained combinations of coefficients for Lorentz violation 
stemming from both the pure-gravity sector \cite{akgrav,lvpn}
and the gravitationally coupled matter sector \cite{akgrav,lvgap,lvgapshort} of the SME are achieved
by reinterpreting the published results of comagnetometer experiments.
One class of coefficients involved in this combination
is difficult to detect, and
with planned improvements in sensitivity,
these experiments 
will be among the proposals \cite{lvgap} 
competitive with the best existing constraints
on coefficients of this class.

In addition to placing constraints on Lorentz violation,
we also explore the possibility of observing
the conventional General-Relativistic gravitomagnetic effect
on intrinsic spin
using anomalous spin-precession experiments.
The gravitoelectric effect
on quantum particles was established
by the experiment of Colella, Overhauser, and Werner (COW) \cite{cow}
and is now observed routinely.
Within General Relativity,
as well as in most alternatives,
the gravitomagnetic effect applies to intrinsic spin
as well as to classical angular momenta,
but establishing this experimentally would 
be of definite interest \cite{ra}.

\section{Basics}
\label{basics}
The relevant theory is a special case 
of the gravitationally coupled SME \cite{akgrav}
investigated in detail in Ref.\ \cite{lvgap}.
In that work,
the theory was treated perturbatively assuming weak gravity,
asymptotically flat spacetime,
slow-moving masses,
and small Lorentz violation.
As such,
the Lorentz-invariant results of General Relativity
and
the flat-spacetime implications of Lorentz violation
are included along with the leading Lorentz-violating modifications to gravity
and can be recovered in the appropriate limit.
Riemann-Cartan spacetime was considered as the geometrical framework
allowing for a nonzero torsion,
an addition warping of spacetime
that can be considered in addition to the curvature of General Relativity.

The matter-sector SME coefficient for Lorentz violation $\bb_\mu$,
plays a key role in the analysis to follow.
In the flat-spacetime limit of the SME,
it enters the action in the form
\beq
\cL_b = - \bar{\ps} \bb_\mu \gamma_5 \gamma^\mu \psi,
\eeq
making its spin coupling evident.

The relativistic hamiltonian relevant for the current discussion
was obtained in Ref.\ \cite{lvgap}.
The result contains a number of effects
having couplings analogous to $\bb_j$.
These contributions
can be written
\beq
H \supset (- \bwt^w_l - \frac{1}{8} T^{\al \be \ga}\ep_{\al \be \ga l} + \quar \prt^j h^{0k} \ep_{jkl}) \ga_5 \ga^0 \ga^l.
\label{ham}
\eeq
The first term here contains the following SME coefficients for Lorentz violation
arising in the matter sector:
\beq
\bwt^w_j= \bb^w_j - \half \ep_{jkl} \Hb^w_{kl} - m^w(\db^w_{0j} - \half \ep_{jkl} \gb^w_{kl0}).
\eeq
The tilde notation denotes combinations of coefficients that arise together
in the nonrelativistic expansion \cite{data}.
In general,
the coefficients are particle-species dependent.
The superscript $w=e,p,n$ indicates
coefficients associated with electrons, protons, or neutrons respectively.
The mass of the relevant particle is denoted $m^w$.
The second term in the hamiltonian
arises due to minimal torsion coupling to fermions.
The presence of this term was exploited in Ref.\ \cite{torsion},
along with nonminimal torsion couplings to fermions,
to place constraints on torsion
using the results
of 
anomalous spin-precession experiments.
The third term in Eq.\ \rf{ham},
is an effective $b_\mu$
containing gravitomagnetic effects.
Here $h_\mn$ is the metric fluctuation defined in terms of
the spacetime metric and the Minkowski metric via the equation
\beq
g_\mn = \et_\mn + h_\mn.
\eeq
Rotating frame effects also enter through this term
when rotating coordinates are used.
At the nonrelativistic level,
the contributions in Eq.\ \rf{ham}
lead to the hamiltonian contributions
\beq
H_{\rm NR} = (- \bwt^w_l - \frac{1}{8} T^{\al \be \ga}\ep_{\al \be \ga l} + \quar \prt^j h^{0k} \ep_{jkl}) \si^l.
\label{nrham}
\eeq

Numerous experiments have constrained $\bwt^w_J$ effects \cite{data,mr,bh,gemmel,expt}.
The capital index here denotes constraints in the Sun-centered frame,
which has been adopted as the standard frame for reporting
sensitivities to SME coefficients in the context of flat-spacetime tests \cite{ssf},
and the concept has been extended to the post-newtonian limit \cite{lvpn}.
At present,
the most sensitive experiment investigating $\bwt^e_J$
is the spin-torsion pendulum at the University of Washington \cite{bh}.
The pendulum bob consists of $\approx 10^{23}$ aligned electron spins
while having negligible magnetic moment.
The most sensitive experiments investigating $\bwt^p_J$
and $\bwt^n_J$ are a He/Xe comagnetometer \cite{gemmel}
and a He/K comagnetometer respectively \cite{mr}.
Comagnetometer experiments exploit the fact
that Lorentz-violation couples to spin rather than magnetic moment.
The colocated magnetometers can then be arranged
such that signals from magnetic fields can be canceled
while achieving impressive sensitivity to Lorentz violation.
The constraints on $\bwt^w_J$
resulting from these experiments are summarized in Table I.
Four orders of magnitude improvement 
over the $\bwt^n_J$ values listed in Table I are expected
in the next generation of comagnetometer experiments \cite{mr}.

\begin{center}
Table I.  Current order of magnitude sensitivities
to $\bwt^w_J$.
\begin{tabular}{c|ccc}
\hline
\hline
           & $w=e$        & $w=p$ & $w=n$ \\
\hline
$\bwt^w_X$ & $10^{-31}$ GeV & $10^{-31}$ GeV & $10^{-32}$ GeV \\
$\bwt^w_Y$ & $10^{-31}$ GeV & $10^{-31}$ GeV & $10^{-32}$ GeV \\
$\bwt^w_Z$ & $10^{-29}$ GeV & - & - \\
\end{tabular}
\end{center}

For comparison with existing and proposed sensitivities to $\bwt_J$,
it is convenient to define an effective coefficient
for the gravitomagnetic effects
entering Eq.\ \rf{nrham}:
\beq
\bgm_l = - \quar \prt^j h^{0k} \ep_{jkl},
\label{bgm}
\eeq
where $h^{0k}$ is understood to contain the
contributions of interest for a given situation.
Note that
$\bgm_l$ corresponds to the usual gravitomagnetic field
in the General Relativity case.

\section{Lorentz-violating effects}
\label{lv}
We first consider Lorentz-violating contributions arising
at second post-newtonian order.
The post-newtonian metric associated with 
the coefficient for Lorentz violation $\sb_\mn$
appearing in the pure-gravity sector of the SME
was obtained in Ref.\ \cite{lvpn}.
A similar analysis was performed
in the context of the gravitationally coupled matter sector
in Ref.\ \cite{lvgap},
obtaining contributions to the post-newtonian metric
from coefficients $\afb_\mu = \ab_\mu -m \eb_\mu$.
For our present interest in lowest-order Lorentz-violating contributions,
the off-diagonal elements the metric fluctuation
appearing in Eq.\ \rf{nrham} can be written
\beq
h_{\rm LV}^{0j} = [\sb^{0j} - \frac{\al}{m} \abs^j] U(x) + [ \sb^{0k} - \frac{\al}{m} \abs^k] U^{jk}(x)
\label{lvmetric}
\eeq
in harmonic gauge.
Here $\al$ represents a nonminimal coupling in the underlying theory
of spontaneous Lorentz-symmetry breaking as discussed
in Ref.\ \cite{lvgap}.
The superscript S on the coefficient $\abs^\mu$
indicates that this is a composite coefficient associated with the particle content
of the source of the gravitational field
defined as
\beq
\abs^\mu = \sum_w N^w \afw^\mu,
\eeq
where $N^w$ is the number of particles of type $w$ contained within the source.
The newtonian potential is denoted $U(x)$,
and 
\beq
U^{jk}(x) = G \int d^3x^\prime \fr{\rh(\vec{x}^\prime,t)R^jR^k}{R^3},
\eeq
where $G$ is Newton's constant,
$R^j = x^j - x^{\prime j}$,
$R = |\vec{x} - \vec{x}^\prime|$,
and $\rh(\vec{x}^\prime,t)$ is mass density.

Using Eqs.\ \rf{bgm} and \rf{lvmetric},
one could in principle calculate the leading Lorentz violating contributions
due to $\sb_\mn$ and $\afb_\mu$ in anomalous spin-precession experiments
for any source.
For the experiments considered here,
Earth is the most relevant source.
At the level of sensitivity available,
it is sufficient to model it as spherically symmetric,
with the local vertical in the lab (3 direction) pointing away from
its center.
Under these conditions
we find the relevant contributions to Eq.\ \rf{bgm} can be written
\beq
\bgmas_j = \half g  \ep_{3kj} \left[\sb^{0k} - \frac{\al}{m_\oplus} \abE^k \right],
\eeq
where $g$ is the local gravitational field of the Earth.
The reader is cautioned that while the analogy between $\bwt_j$
and $\bgmas_j$ exists in the laboratory frame,
there is not typically a direct match between the relevant 
coefficients for Lorentz violation
in the Sun-centered frame due to the nontrivial dependence
of $\bgmas_j$ on the coefficients $\sb^{0k} - \frac{\al}{m_\oplus} \abE^k$.
For example,
for experiments performed on the surface of the Earth,
the time-dependence of $\bwt_j$ 
can be displayed explicitly as
\bea
\nonumber
\bwt_1 &=& \bwt_X \cos \ch \cos \om T + \bwt_Y \cos \ch \sin \om T - \bwt_Z \sin \ch\\
\nonumber
\bwt_2 &=& -\bwt_X \sin \om T + \bwt_Y \cos \om T\\
\bwt_3 &=& \bwt_X \sin \ch \cos \om T + \bwt_Y \sin \ch \sin \om T + \bwt_Z \cos \ch,
\label{lvtdep}
\eea
while for $\bgmas_j$ the time dependence takes the form
\bea
\nonumber
\bgmas_1 &=&  \half g [\sb^{TX} - \frac{\al}{m_\oplus} \abE^X] \sin \om T\\
\nonumber
\pt{\bgmas_1 =} & &
- \half g [\sb^{TY} - \frac{\al}{m_\oplus} \abE^Y] \cos \om T\\
\nonumber
\bgmas_2 &=& \half g [\sb^{TX} - \frac{\al}{m_\oplus} \abE^X] \cos \ch \cos \om T\\
\nonumber
\pt{\bgmas_2 =} & &
+ \half g [\sb^{TY} - \frac{\al}{m_\oplus} \abE^Y] \cos \ch \sin \om T\\
\nonumber
\pt{\bgmas_2 =} & &
- \half g [\sb^{TZ} - \frac{\al}{m_\oplus} \abE^Z] \sin \ch\\
\bgmas_3 &=& 0.
\label{bgmaslong}
\eea

In spite of this subtlety,
sufficient information exists to place constraints using published limits
on $\bwt_J$.
Note that 
the most sensitive investigations of $\bwt_J$ may be applied
without regard to flavor,
since the gravitomagnetic effects considered here 
are independent of test-body flavor.

Ignoring flavor,
the best sensitivities to $\bwt_{X,Y}$ at present
come from the He/K comagnetometer experiment \cite{mr}.
In that work,
measurements were initially made of
a magnetic field-like parameter $\be^N$
with hamiltonian contribution
$H \supset -\mu_{^3He} \be^N_i \si^N_i$,
where the superscript $N$ denotes quantities associated with the nucleus,
and $\mu_{^3He}$ is the magnetic moment of the $^3He$
nucleus.
Measurements of
\bea
\nonumber
\be^N_X &=& (-0.020 \pm 0.040) \,\, \rm{fT}\\
\be^N_Y &=& (0.061 \pm 0.051) \,\, \rm{fT}
\eea
were obtained based on investigation of the 1 direction,
and consideration of the 2 direction yielded \cite{jb}
\bea
\nonumber
\be^N_X &=& (0.011 \pm 0.024) \,\, \rm{fT}\\
\be^N_Y &=& (0.025 \pm 0.022) \,\, \rm{fT}.
\eea
These results were then combined and interpreted
as constraints on $\bwt^n_{X,Y}$ in Ref.\ \cite{mr}.
Here they can be used along with Eq.\ \rf{bgmaslong}
to obtain the following results:
\bea
\nonumber
\sb^{TX} - \frac{\al}{m_\oplus} \abE^X &=& (0.24 \pm 0.15)\\
\sb^{TY} - \frac{\al}{m_\oplus} \abE^Y &=& (0.02 \pm 0.13).
\eea
Here all uncertainties are 1 sigma.

With the expected improvements in the He/K comagnetometer experiment \cite{mr},
sensitivities at the level of $10^{-4}$ should be attained.
The composite $\afb_J$ coefficients
for Earth appearing in these results
can be expanded as
\beq
\frac{1}{m_\oplus} \abE^J =
(0.54{\rm GeV^{-1}})[\afe^J + \afp^J + \afn^J],
\eeq
using $N^e=N^p \approx N^n = 1.8 \times 10^{51}$ for Earth \cite{earth}.
This implies sensitivity competitive with 
the maximum reach achieved to date on $\abn_J$
and $\ebn_J$ \cite{data}
will be attained.
Moreover,
the constraints are on different combinations
of coefficients than those involved in the existing tests \cite{lvgap,oa}.
Thus combining results
would yield additional independent sensitivities
to Lorentz violation.

We note in passing that a full investigation
of Lorentz violating couplings to spin
in the presence of gravity may yield additional sensitivities
to matter-sector coefficients
associated with the test body;
however, such an investigation
is beyond the scope of the present work.

\section{Conventional gravitomagnetic effects}
\label{conventional}
The conventional General-Relativistic gravitomagnetic effect
is more challenging to detect in anomalous spin-precession experiments
for several reasons.
However,
it may in principle soon fall within the reach of these experiments.
Presently,
the largest gravitomagnetic effect on spins in the lab
is that of Earth.
Thus it is interesting to consider the effective $b_\mu$
due to the conventional gravitomagnetic field of Earth
by inserting the appropriate $h^{0j}$
into Eq.\ \rf{bgm}.
By modeling Earth as a uniformly rotating sphere of radius $R_\oplus$,
the angular momentum of Earth due to its 
rotation on its axis at angular speed $\om$ is
$\vec J_\oplus = \frac{2}{5}m R_\oplus^2 \om \hat{Z}$,
and the relevant quantity for insertion in Eq.\ \rf{bgm}
can be written
\beq
h_\oplus^{0j} = \fr{2G}{r^3} \ep_{jkl} (J_\oplus)_k x_l.
\eeq
This yields
\bea
\nonumber
\bgmE_1 &=& - \fr{G}{2r^3} J_\oplus \sin \ch\\
\nonumber
\bgmE_2 &=& 0\\
\bgmE_3 &=& - \fr{G}{r^3} J_\oplus \cos \ch
\label{bgmE}
\eea
for the explicit form of $\bgmE_j$,
which is of order $10^{-38}$GeV
on the surface of the Earth.

Although the effect lies within about 2 orders of magnitude
of the expected sensitivity of the next generation of experiments,
an observation of the effect would be challenging
even if the required sensitivity were reached.
At present,
the relevant experiments
are performed in the rotating reference frame of the Earth
where an effective $b_\mu$ with components
\bea
\nonumber
\bnif_1 &=& - \half \om \sin \ch\\
\nonumber
\bnif_2 &=& 0\\
\bnif_3 &=& \half \om \cos \ch
\label{bnif}
\eea
arises due to noninertial frame gyroscopic effects.
Note that the effect in Eq.\ \rf{bgmE}
has components in the same direction as the larger contributions in Eq.\ \rf{bnif}.
The effects could in principle be distinguished by their different dependencies on $r$,
but doing so would likely be challenging in practice.
Another approach would be to perform an anomalous spin-precession experiment in space,
as has been suggested in the context of detecting curvature components \cite{mohanty}.
Here a frame could be chosen that breaks the symmetry between
Eq.\ \rf{bgmE} and Eq.\ \rf{bnif}
and/or the $r$ dependence of Eq.\ \rf{bgmE} could be better exploited.

Though their size lies well below the expected sensitivity of anomalous spin-precession experiments
in the near term,
it is worth commenting on conventional gravitomagnetic effects arising from the Sun
and rotating masses in the laboratory.
Following the same procedure leading to Eq.\ \rf{bgmE},
one can obtain an effective $b_\mu$ for the gravitomagnetic effect
of the Sun, which can be written
\beq
\bgmS_J = - \half G \left[\fr{3 (J_\odot)_K x_K x_J}{r^5} - \fr{(J_\odot)_J}{r^3} \right].
\label{bgmS}
\eeq
Although about 6 orders of magnitude smaller than $\bgmE_j$ contributions,
$\bgmS_J$ has time dependence in the lab that is similar to Lorentz violation
and decoupled from the noninertial frame effects.
If the rotation axis of the Sun were exactly perpendicular to the ecliptic,
the conventional gravitomagnetic effects due to the Sun
in Earth-based anomalous spin-precession experiments could be obtained
by inserting 
the simple constant expressions
$\bgmS_Y \approx - \fr{G}{2 R_{\rm E S}^3} J_\odot \sin \et$
and
$\bgmS_Z \approx \fr{G}{2 R_{\rm E S}^3} J_\odot \cos \et$
into Eq.\ \rf{lvtdep}.
Here $R_{\rm E S}$ is the Earth-Sun distance,
$\et$ is the angle of the $XY$ plane relative to the ecliptic,
and a circular orbit has been assumed.

A more detained analysis taking into account the inclination
of the rotation axis of the Sun at an angle $i=7.25^o$ 
relative to the normal to the ecliptic
yields
\bea
\nonumber
& & \bgmS_X = \fr{- G J_\odot}{4 R_{\rm E S}^3} \sin i 
\pt{\fr{- G J_\odot}{4 R_{\rm E S}^3} \cos \et \sin i 19999999999}\\
\nonumber
& &
\pt{b_{gm}} \times 
 [\cos \Th (3 \cos 2 \Om T -1) + 3 \sin \Th \sin 2 \Om T]\\
\nonumber
& & \bgmS_Y = \fr{- G J_\odot}{2 R_{\rm E S}^3} \sin \et \cos i
+ \fr{G J_\odot}{4 R_{\rm E S}^3} \cos \et \sin i \\
\nonumber
& &
\pt{b_{gm}} \times [\sin \Th (3 \cos 2 \Om T -1) + 3 \cos \Th \sin 2 \Om T]\\
\nonumber
& & \bgmS_Z = \fr{G J_\odot}{2 R_{\rm E S}^3} \cos \et \cos i
+ \fr{G J_\odot}{4 R_{\rm E S}^3} \sin \et \sin i \\
& &
\pt{b_{gm}} \times [\sin \Th (3 \cos 2 \Om T -1) + 3 \cos \Th \sin 2 \Om T].
\eea
Here $\Om$ is the angular speed of Earth along its orbit,
and $\Th \approx 14^o$ is the present angular distance along the ecliptic
from the projection of the solar spin
to the vernal equinox \cite{smart}.
The interesting time dependence at the annual frequency arises from the first term in
Eq.\ \rf{bgmS}
and is due to the motion of the experiment
through the solar gravitomagnetic field.
These annual variations
along with the radial dependence
could help distinguish these effects
from Lorentz violation.

The possibility of large angular momenta in the lab 
has also improved in recent years due to interest in high rotation-rate
flywheels for energy storage.  
Though the effective $b_\mu$ that could be generated by such devices
in anomalous spin-precession experiments is at least 8 orders of magnitude smaller
than $\bgmE_j$, the effect could easily be controlled and modulated
allowing easy separation from other effects.
From the stand point of SME-based tests of Lorentz symmetry,
these systems would also offer knowledge and control
of the source composition.
There are also Lorentz-violating contributions
to $h^{0j}$ that lie at the same post-newtonian order 
as the conventional General-Relativistic gravitomagnetic effects,
and an investigation of their implications may become interesting if sufficient
sensitivity is attained in anomalous spin-precession experiments.

\section{Summary}
\label{summary}
In this work we have seen the unexpected result
that experiments searching for anomalous precessions of intrinsic spins,
which were designed to test Lorentz symmetry in a nongravitational context,
have the ability to place new constraints on Lorentz-violating effects
via gravitomagnetism.
New constraints are placed on the SME coefficients
for Lorentz violation $\sb_\mn$ and $\afb_\mu$
using existing data,
and the next generation of these experiments
is expected to yield sensitivities
competitive with the best existing sensitivities
to coefficients of this type.
We also find that these experiments may one day
observe the conventional gravitomagnetic effect
on intrinsic spin.
This demonstrates the continuing impact
of Lorentz violation searches
on the experimental investigation
of fundamental physics.

\end{document}